\documentclass[english]{aastex}
\usepackage{amssymb}
\usepackage{graphicx}

\makeatletter

\slugcomment{}
\shorttitle{Constraining IMBHs in Globular Clusters}
\shortauthors{Umbreit \& Rasio}

\makeatother

\usepackage{babel}
\begin{document}

\title{Constraining Intermediate-Mass Black Holes in Globular Clusters }

\author{Stefan Umbreit}

\affil{Center for Interdisciplinary Exploration and Research in Astrophysics
(CIERA) \& Dept. of Physics and Astronomy, Northwestern University,
2145 Sheridan Rd, Evanston, IL 60208, USA}

\email{s-umbreit@northwestern.edu}

\and{}

\author{Frederic A. Rasio}

\affil{Center for Interdisciplinary Exploration and Research in Astrophysics
(CIERA) \& Dept. of Physics and Astronomy, Northwestern University,
2145 Sheridan Rd, Evanston, IL 60208, USA}

\email{rasio@northwestern.edu}
\begin{abstract}
Decades after the first predictions of intermediate-mass black holes
(IMBHs) in globular clusters (GCs) there is still no unambiguous observational
evidence for their existence. The most promising signatures for IMBHs
are found in the cores of GCs, where the evidence now comes from the
stellar velocity distribution, the surface density profile, and, for
very deep observations, the mass-segregation profile near the cluster
center. However, interpretation of the data, and, in particular, constraints
on central IMBH masses, require the use of detailed cluster dynamical
models. Here we present results from Monte Carlo cluster simulations
of GCs that harbor IMBHs. As an example of application, we compare
velocity dispersion, surface brightness and mass-segregation profiles
with observations of the GC M10, and constrain the mass of a possible
central IMBH in this cluster. We find that, although M10 does not
seem to possess a cuspy surface density profile, the presence of an
IMBH with a mass up to $0.75\%$ of the total cluster mass, corresponding
to about $600\,{\rm M}_{\odot}$, cannot be excluded. This is also
in agreement with the surface brightness profile, although we find
it to be less constraining, as it is dominated by the light of giants,
causing it to fluctuate significantly. We also find that the mass-segregation
profile cannot be used to discriminate between models with and without
IMBH. The reason is that M10 is not yet dynamically evolved enough
for the quenching of mass segregation to take effect. Finally, detecting
a velocity dispersion cusp in clusters with central densities as low
as in M10 is extremely challenging, and has to rely on only $20-40$
bright stars. It is only when stars with masses down to $0.3\,{\rm M_{\odot}}$
are included that the velocity cusp is sampled close enough to the
IMBH for a significant increase above the core velocity dispersion
to become detectable.
\end{abstract}


\section{Introduction}

It has been known for a long time that cusps in the velocity dispersion
or density profiles could provide strong indication for the presence
of an intermediate-mass black hole (IMBH) at the center of a star
cluster \citep[see, e.g.,][for a short review]{2012ApJ...750...31U}.
The existence of these black holes, with masses intermediate between
stellar ($M_{BH}\lesssim20M_{\odot}$), and supermassive ($M_{BH}\gtrsim10^{6}M_{\odot}$),
could not be established by observations up until recently, although
IMBHs were discussed by theorists already more than 30 years ago \citep[see, e.g.,][]{1970ApJ...160..443W}.
In contrast to indirect evidence from, e.g., so-called ultra-luminous
X-ray sources (ULXs), sources with luminosities that exceed what can
be produced by a stellar mass BH accreting at the Eddington limit,
measurements of the cusp slopes allow a more direct determination
of the mass of an IMBH. 

However, such cusps might not be easily detectable in real star clusters.
On the one hand, cusps in the surface brightness profile (SBP) are
expected to be rather shallow, making them difficult to distinguish
from standard King models \citep{2005ApJ...620..238B}. On the other
hand, measurements of cusps in velocity dispersion profiles, though
steeper, have to rely on only a relatively small number of stars for
typical globular clusters \citep{2004ApJ...613.1133B}. This especially
applies to the old Milky Way globular clusters (hereafter GCs), where
the cluster centers are most likely dominated by dark stellar remnants,
reducing the number of observable bright stars in the cusp. Based
on SBP measurements by \citet{2006AJ....132..447N}, 9 candidate GCs
with inner SBP power-law slopes in the range $-0.1$ to $-0.3$, indicative
of an IMBH, have been identified so far \citep{2005ApJ...620..238B}.
These slopes represent, however, only tentative evidence as their
error bars, based on photometric and statistical errors, are rather
large, ranging from 50-100\% of the slope value. In addition, from
$N$-body and Monte Carlo simulations it has been found that these
slopes are rather time variable such that, e.g., the inner SBP of
a cluster with IMBH could even be completely flat for some brief period
of time \citep{2012ApJ...750...31U,2011ApJ...743...52N,2010ApJ...720L.179V,2010ApJ...708.1598T}.

Another measure for the mass of an IMBH comes from the size of the
cluster core, with more massive IMBHs producing larger cores as measured
by the core-to-half light ratio \citep[e.g.,][]{2007PASJ...59L..11H,1980ApJ...239..685M,1977ApJ...217..281S}.
As shown by \citet{2007MNRAS.381..103M}, the size of the core is
related to the inner SBP cusp slope such that clusters with larger
slopes have also lower concentrations if they contain an IMBH in their
center. Thus, the structure of GCs together with their inner SBP slopes
should in principle have the potential to lead to stronger constraints
by placing upper \emph{and }lower limits on the IMBH mass. Based on
literature values for the inner slopes and concentrations, there are
2 clusters (from the previous list of 9 IMBH candidate clusters of
\citet{2005ApJ...620..238B}) whose concentrations are inconsistent
with their inner SBP slopes, leaving 7 IMBH cluster candidates. However,
as with the inner slopes of the SBP, the determination of the concentration
is plagued with considerable uncertainties. This is mainly because
the escape of unbound stars is delayed \citep{2000MNRAS.318..753F},
and thus the escaping stars may contribute significantly to the outer
cluster light. The escaping stars, if not considered, can lead to
a much larger apparent tidal cut-off radius, and, thus, to a significantly
larger estimate of the concentration parameter \citep{2010ApJ...708.1598T}.
For instance, in \citet{2012ApJ...750...31U}, we have shown that,
although the concentration of the cluster NGC~5694, quoted as $1.8$
in the Harris catalog, is too large for the inner SBP slope of $-0.19$,
it is nevertheless still consistent with the presence of an IMBH if
the flattening of the outer SBP can be attributed to the stellar background.

A less sensitive measure of the mass, but more sensitive to the \emph{presence}
of an IMBH, recently proposed by \citet{2008ApJ...686..303G}, is
based on the average mass profile of main-sequence stars normalized
to its value at the half-mass radius. Their direct $N$-body simulations
show that an IMBH changes the mass-segregation profile of the cluster
such that it effectively counterbalances the tendency of massive stars
to concentrate towards the center, similar to what has been reported
earlier by \citet{2004ApJ...613.1143B}. Both \citet{2004ApJ...613.1143B}
and \citet{2008ApJ...686..303G} suggest that the reason for the so-called
``quenching'' of mass segregation is strong binary interactions
between the massive stars sinking towards the center and any companion
stars bound to the IMBH. Furthermore, a similar mechanism might be
responsible for the somewhat weaker quenching of mass segregation
in clusters with primordial binaries as in both cases strong binary
interactions are involved. 

The mass segregation signature does, however, require that the cluster
has had enough time to settle down to a dynamically relaxed state,
which happens over several half-mass relaxation times. Indeed, many
old Milky Way GCs might be in such a relaxed state, as their estimated
half-mass relaxation times are an order of magnitude shorter than
their age \citep[e.g.,][]{2005ApJS..161..304M}. As shown in many
studies, relaxed stellar systems evolve towards a self-similar configuration,
and the cluster structure is then mostly determined by heating processes
in the cluster core, such as interactions with primordial binaries
\citep{2003ApJ...593..772F,2007ApJ...658.1047F}, or with an IMBH
\citep{2004ApJ...613.1143B,2005ApJ...620..238B,2007PASJ...59L..11H,2007MNRAS.374..857T},
the formation of three-body binaries \citep{1984MNRAS.208..493B},
and stellar collisions \citep{2008IAUS..246..151C}. From a modeling
perspective, this has the advantage that the parameter space of initial
cluster conditions to explore for a given observed GC is greatly reduced,
as the structure of relaxed stellar systems is rather independent
from the conditions at the time of their formation.

However, \citet{2012ApJ...750...31U} demonstrate that dynamic age
estimates based only on the current state of clusters are very unreliable,
for two reasons. First, the half-mass relaxation time is generally
time dependent, so there is a dependence of the dynamic age on the
previous evolutionary history of the cluster \citep[see also][]{2007MNRAS.379...93H}.
Second, given a final state, the dynamic age can differ significantly
depending on the initial cluster size with respect to the tidal boundary.
For instance, when a cluster significantly underfills its Roche lobe
it expands freely, increasing its relaxation time, whereas for tidally
filling clusters the relaxation time can decrease as the cluster cannot
expand beyond the tidal radius while its core is contracting \citep[see, e.g.,][]{2007MNRAS.379...93H}.
As an added difficulty, it may be hard to decide in which of these
categories the evolution of a particular observed cluster falls, as
the tidal field the cluster experienced may have been strongly time-dependent
and could require extensive analysis to constrain \citep[e.g.,][]{2011MNRAS.414.1339K,2006ApJ...652.1150A}.
As a consequence of the uncertainties in the dynamic age estimates,
it is not clear\emph{ a} \emph{priori} whether a cluster is in the
fully relaxed state or, possibly, still in its core contraction phase,
which may have consequences for the observability of IMBH signatures.

From this discussion it becomes clear that meaningful constraints
on the mass of a possible IMBH at the center of an observed GC require
extensive analysis of detailed, realistic evolutionary cluster models,
which are then compared to observations. In this paper we extend our
analysis in \citet{2012ApJ...750...31U} and consider, in addition
to the SBP, kinematic and mass segregation signatures. As an example
of direct comparison to observations we focus on the cluster M10 (NGC~6254).
This cluster is especially suited for such a comparison given its
close proximity to the sun ($\approx4\,{\rm kpc}$) and multitude
of available observational data \citep{2006AJ....132..447N,2010ApJ...713..194B,2011ApJ...743...11D}.
In addition to considering more observables, we now also consider
clusters that fill their Roche lobes and are significantly influenced
by tidal stripping.

Our paper is organized as follows. In Section 2 we briefly present
the cluster Monte Carlo method, and describe initial conditions for
the simulations and observations of M10. In Section 3 we present surface
density, surface brightness, average mass profile, and velocity dispersion
profiles from our models with and without central IMBH, and determine
the maximum mass a hypothetical IMBH could possess to be still compatible
with observations. We conclude in Section 4.

\section{Initial Conditions and Observations}

\subsection{Initial Conditions}

We carried out a large parameter study to model GCs with and without
IMBH using our Cluster Monte Carlo code \citep[CMC;][]{2012ApJ...750...31U,2010ApJ...719..915C,2007ApJ...658.1047F,2003ApJ...593..772F,2001ApJ...550..691J,2000ApJ...540..969J}.
As with direct $N$-body codes \citep[e.g.,][]{2003gnbs.book.....A,2001MNRAS.321..199P},
CMC uses a star-by-star, discrete representation of the cluster, and
the code is now able to treat all relevant processes such as stellar
evolution \citep{2010ApJ...719..915C}, strong interactions between
stars and binaries \citep{2007ApJ...658.1047F}, as well as the dynamics
of a central massive BH, including an advanced treatment of the loss
cone physics needed to accurately estimate the rate of cluster heating
\citep{2012ApJ...750...31U}. The dynamical evolution of the cluster
is computed on the relaxation timescale (i.e., the timestep is a fraction
of the relaxation time, $\sim10^{9}\,{\rm yr}$ for a typical GC,
rather than the much shorter dynamical time, $\sim10^{6}\,{\rm yr}$)
which enables us to calculate the evolution of massive GCs in a relatively
short time (typically less than a week on a modern workstation).

Our study consists of approximately 500 model calculations varying
IMBH mass, initial number of stars, cluster concentration, and distance
from the Galactic center. The positions and velocities of the stars
are initially distributed according to King models, and their masses
are chosen according to the Kroupa IMF \citep{2001MNRAS.322..231K}
in the range from $0.1$ to $100\,{\rm M_{\odot}}$. The initial virial
radius for all clusters has been selected such that the cut-off radius
of the King models coincides with the cluster's Roche, or tidal, radius,
$r_{t}$, in an external tidal Galactic field, given by 
\begin{equation}
r_{t}=\left(\frac{GM_{c}}{2V_{G}^{2}}\right)^{1/3}R_{G}^{2/3}\label{eq:rt}
\end{equation}
where $M_{c}$ is the total cluster mass, $V_{G}$ and $R_{G}$ the
Galactic circular velocity and galactocentric distance, respectively.
For all runs we set $V_{G}=220\,{\rm km\, s^{-1}}$, the standard
value for the Milky Way. As in our previous study we included $10\%$
hard binaries in some of our models, given their potential influence
on the SBP \citep{2010ApJ...720L.179V} and mass segregation profile
\citep{2008ApJ...686..303G}. Each star was tested for entry into
the loss-cone, the region of angular momentum and energy space where
the periapse of the stellar orbit is smaller than the tidal disruption
radius of the star around the IMBH. See \citet{2012ApJ...750...31U}
for a detailed description. The metallicity, $Z$, for the stars was
set to $Z=1\times10^{-3}$, and all models are evolved for $12\,{\rm Gyr}$.
We calculated the surface brightness and surface density profile by
converting the stellar radius and bolometric luminosity for each star
obtained from the CMC stellar evolution module BSE to V-band luminosity
using the standard stellar library in \citet{1998A&AS..130...65L}.
Then, certain selection criteria are applied, the stellar positions
projected onto the sky, and the stars binned, all in correspondence
with the relevant observations (see next section). Table \ref{tab:Parameter-ranges-explored}
summarizes the parameter ranges explored.

\begin{table}
\begin{tabular}{cc}
Parameter & Range\tabularnewline
\hline 
$N$ & $2.8-8\times10^{5}$\tabularnewline
$W_{0}$ & $5-7$\tabularnewline
$M_{BH}$ & $300-2000\,{\rm M_{\odot}}$\tabularnewline
$R_{G}$ & $0.9-4.1\,{\rm kpc}$\tabularnewline
$Z$ & $0.001$\tabularnewline
\end{tabular}

\caption{\label{tab:Parameter-ranges-explored}Parameter ranges explored. All
models were tidally filling, and their half-mass radii, thus, given
by $r_{t}$ from Equation \ref{eq:rt} and $W_{0}$, range approximately
from $2-11\,{\rm pc}$. }
\end{table}

\subsection{Data}

We use the Milky Way GC M10 (NGC~6254) as an example for constraining
the mass of a hypothetical IMBH. Here we work with the SBP from \citet{2006AJ....132..447N},
taken with the WFPC2 camera on the \emph{Hubble Space Telescope }for
the inner $1.7\,{\rm pc}$, and for the outer region with the corresponding
data from the Trager catalog \citep{1995AJ....109..218T}. We compare
our models by, first, converting the absolute V-band luminosities
of each star, we obtained as described in the previous section, to
apparent magnitudes using a distance of M10 to the sun of $4.4\,{\rm kpc}$
\citep{1996AJ....112.1487H}. The stars were then projected onto the
sky and binned using similar bin sizes as in \citet{2006AJ....132..447N}
and \citet{1995AJ....109..218T}. In order to minimize the large fluctuations
caused by bright giants, we impose a brightness cut-off, and only
consider objects with an absolute V-band magnitude fainter than that.
The cut-off is chosen as a compromise between a smooth profile and
little bias in the profile shape. 

Thanks to the ``ACS Survey of Galactic Globular Clusters'' (GO-10775;
PI: A. Sarajedini) there are also deep, high-resolution observations
available, with excellent photometry of main-sequence stars with masses
down to $0.3\,{\rm M_{\odot}}$ \citep{2010ApJ...713..194B}. The
high resolution makes it possible to construct a sufficiently detailed
star count profile, or surface density profile (SDP), for the central,
denser regions of M10. For the star count profile stars were binned
in concentric annuli around the center, as determined in \citet{2010ApJ...713..194B}.
The bin widths ensure that the individual surface density values are
based on at least 100 stars with the exception of the innermost bin,
which contains only 38 stars. Only stars with an apparent V-band magnitude
brighter than $19$ are considered because, with a completeness fraction
of $\approx90\%$ \citep{2010ApJ...713..194B}, they do not suffer
significantly from crowding throughout the entire cluster. At a distance
of $4.4\,{\rm kpc}$ from the sun and a metallicity of $Z=0.001$
this magnitude corresponds to a main-sequence star with mass $\approx0.7\,{\rm M_{\odot}}$.

\begin{figure}
\begin{tabular}{cc}
\includegraphics[width=0.45\columnwidth]{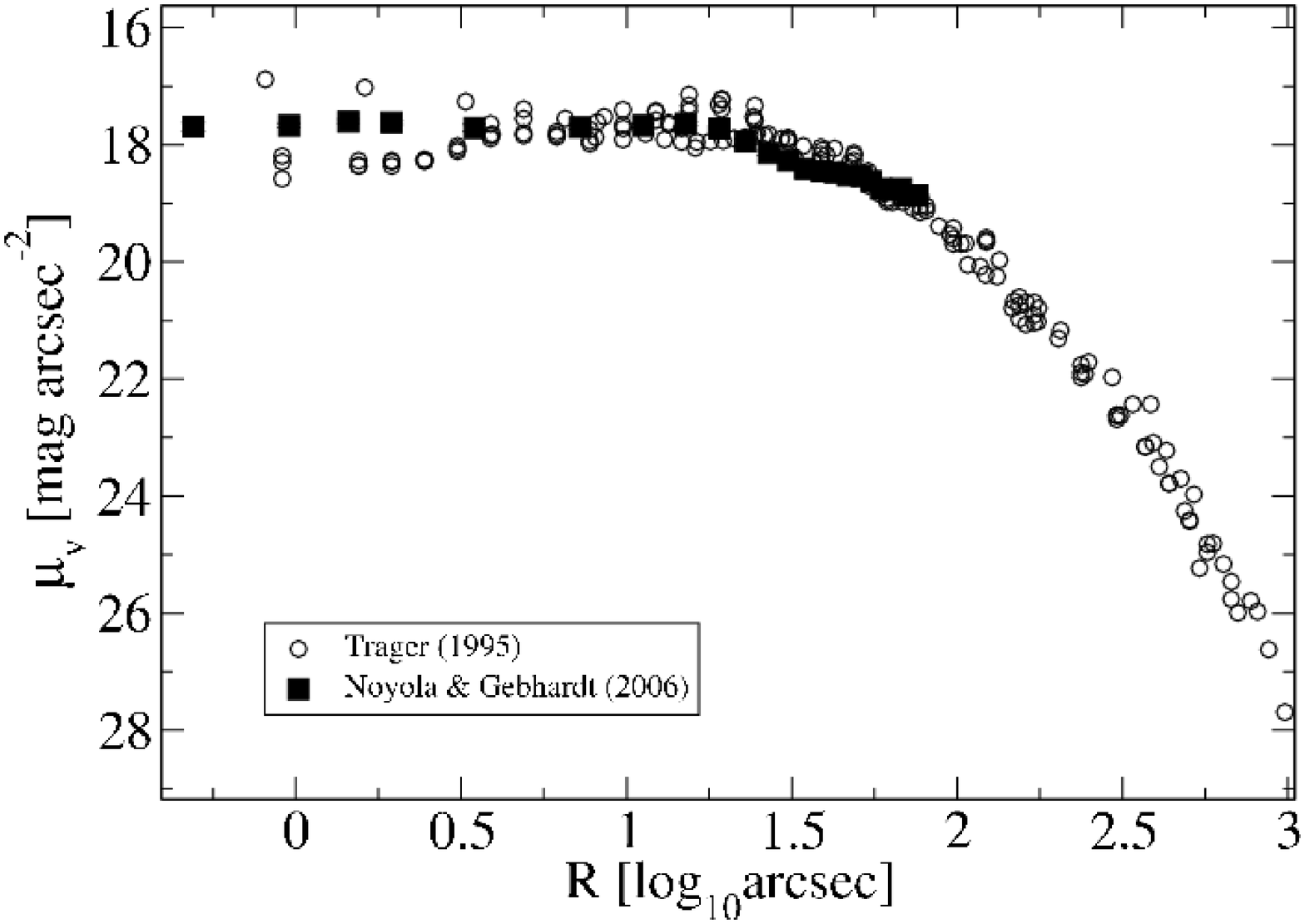} & \includegraphics[clip,width=0.45\columnwidth]{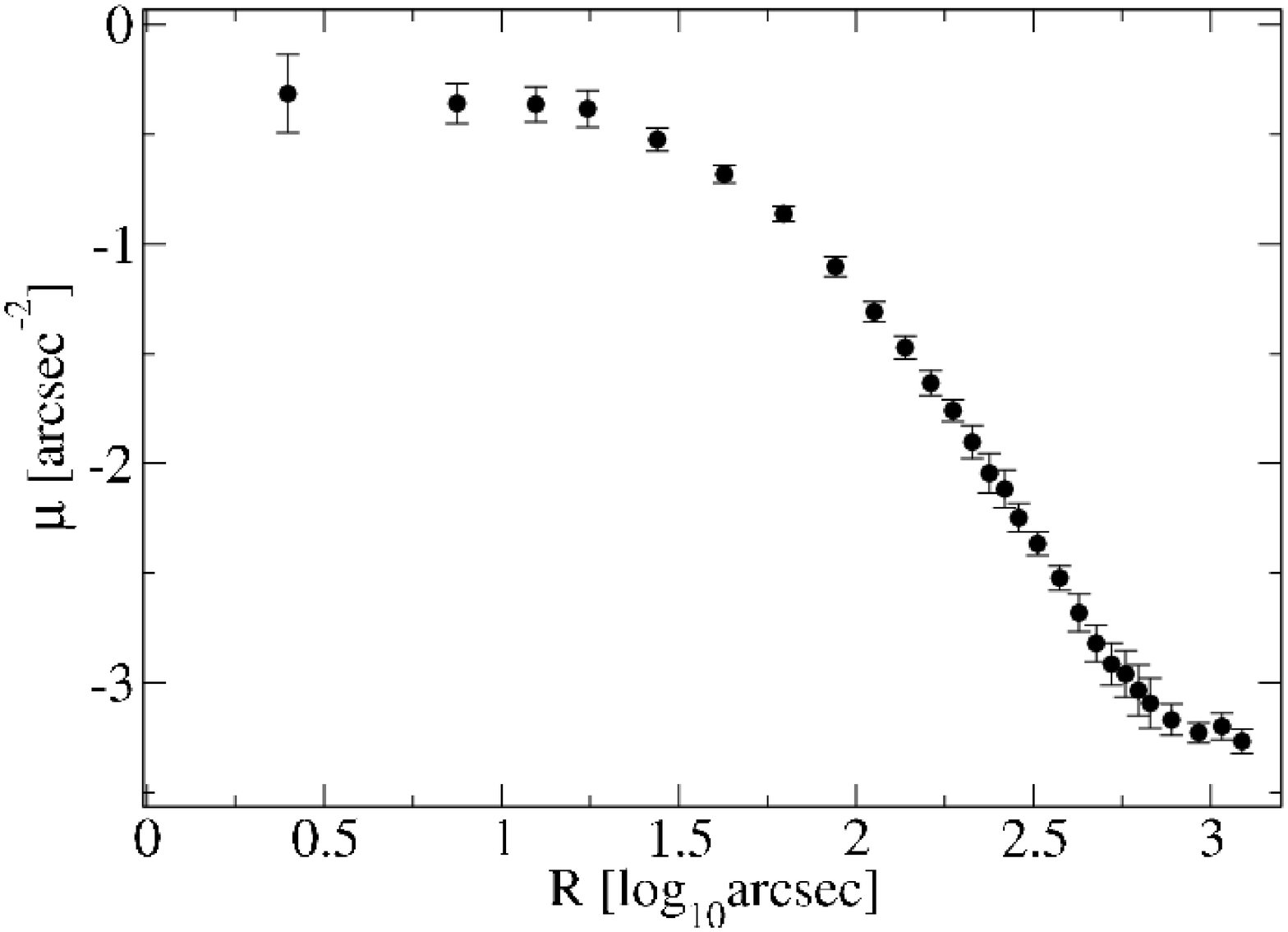}\tabularnewline
\end{tabular}\caption{\label{fig:surface-brightness-density-obs}Surface brightness profile
(left panel) and star count profile (right panel) of M10. The surface
brightness profile combines data from \citet{2006AJ....132..447N}
and \citet{1995AJ....109..218T}. The error bars of the star count
profiles represent the Poisson error.}
\end{figure}

Figure \ref{fig:surface-brightness-density-obs} shows the observed
SBPs from \citet{2006AJ....132..447N} and the star count profile
for M10. As can be seen,
M10 does not posses a notable negative inner SBP or SDP slope, and
\citet{2006AJ....132..447N} derive even a slightly positive value
of $0.05$, all not indicative of the presence of an IMBH. On the
other hand, as we have shown previously \citep{2012ApJ...750...31U},
the SBP slopes are highly variable, in particular for clusters with
IMBH, and therefor, the possibility remains that M10's profile might
still be consistent with the presence of an IMBH at its center.

A simplification in choosing our initial conditions is that we only
consider tidally filling clusters. However, the observed SBP cut-off
radius of M10 is about $20\:{\rm pc}$, while the tidal radius of
a cluster with mass $M_{c}\approx1.5\times10^{5}\,{\rm M_{\odot}}$
\citep{2005ApJS..161..304M} at a galactocentric radius of $4.6\,{\rm kpc}$
\citep{1996AJ....112.1487H} is $r_{t}\approx50\,{\rm pc}$. We immediately
conclude that M10 is currently underfilling its Roche lobe significantly.
This filling factor does not change substantially at periapse passage
for an orbital eccentricity of $0.19$, determined by \citet{1999AJ....117.1792D}
based on an axisymmetric model for the Galactic potential. However,
when the existence of bar structures in the inner $\approx4\,{\rm kpc}$
is approximately accounted for, \citet{2006ApJ...652.1150A} demonstrate
that M10 might have experienced a much stronger tidal field at perigalacticon
distances sometimes as small as $1\ {\rm kpc}$. The large radial
excursions are a direct consequence of the gravitational interactions
between the cluster and the rotating bar potential, sometimes leading
to the trapping of clusters in resonances and irregular orbits \citep{2006ApJ...652.1150A}.
In our study we avoid to model the complex orbit of M10, and, instead,
evolve our clusters at a fixed galactocentric radius, representing
an average tidal field.

\section{Results}

\subsection{Surface Density Profile}

\begin{figure}
\begin{tabular}{cc}
\includegraphics[clip,width=0.45\columnwidth]{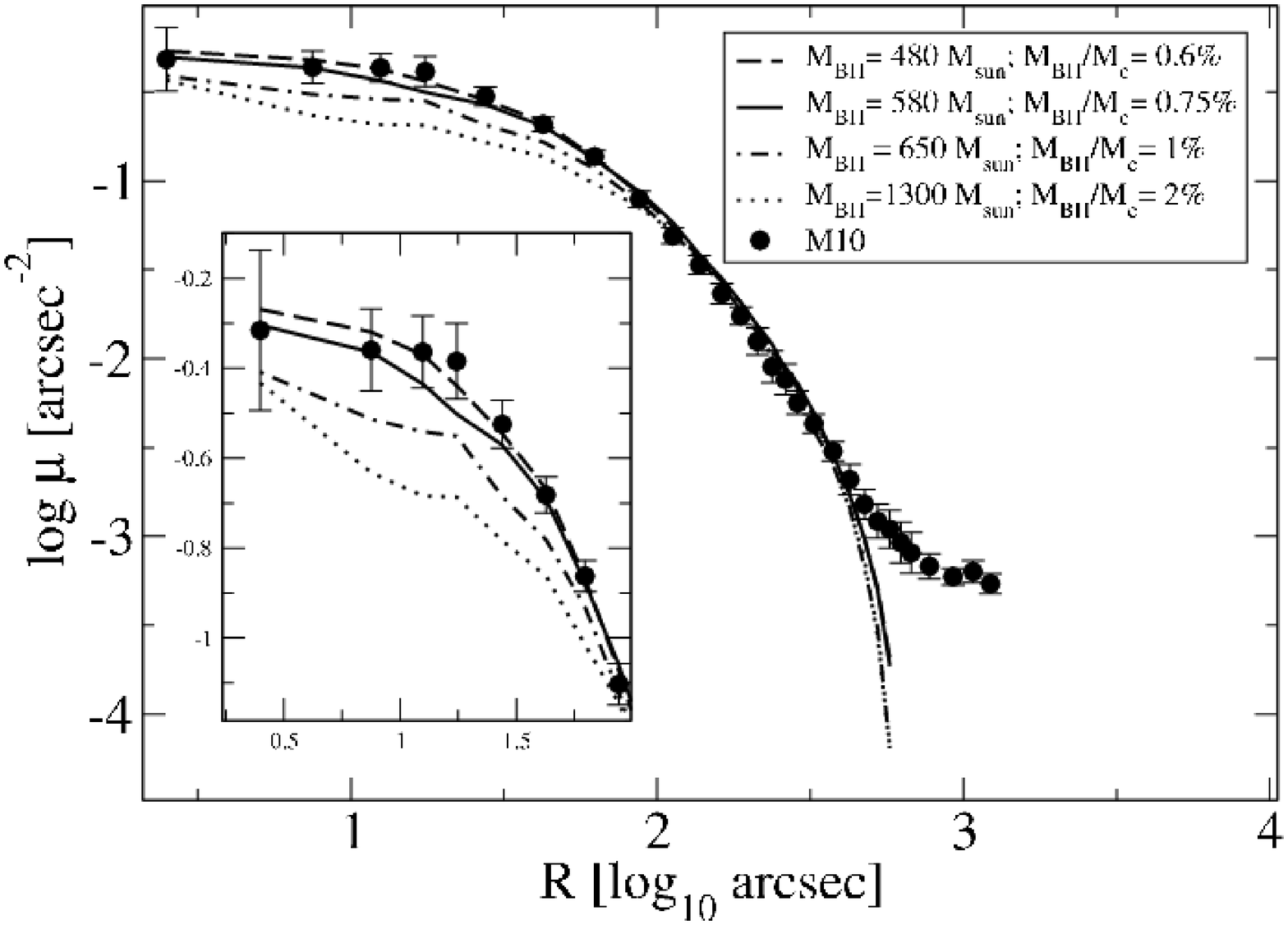} & \includegraphics[clip,width=0.45\columnwidth]{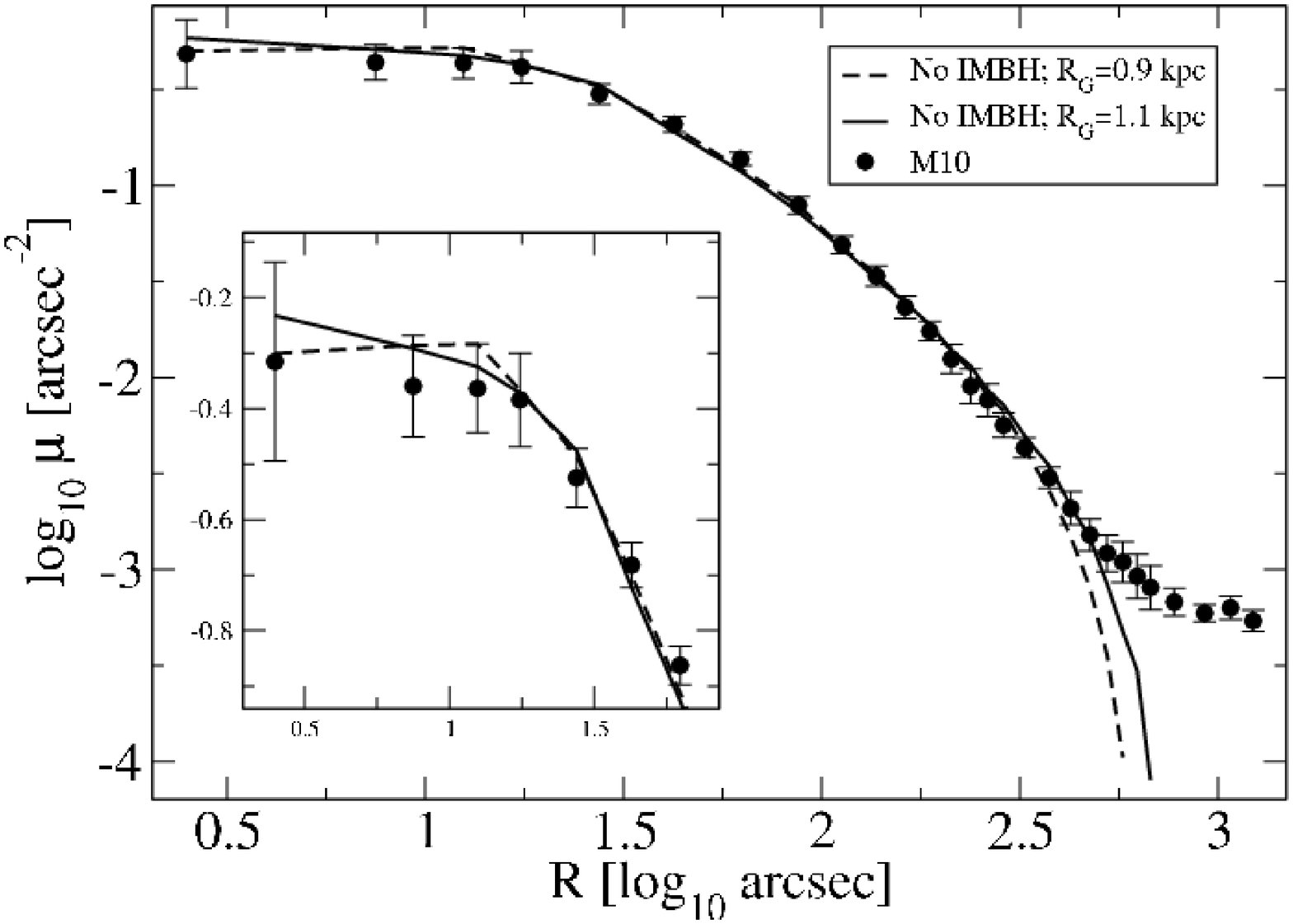}\tabularnewline
\end{tabular}

\caption{\label{fig:Surface-density-profiles}Surface density profiles from
our best-fit model with (left panel) and without (right panel) a central
IMBH for M10 (filled circles). Left: models with different IMBH masses
at galactocentric radius of $0.9\,{\rm kpc}$. The maximum IMBH mass
for which a reasonable match to the data can be obtained is $580\,{\rm M_{\odot}}$.
Right: Models without IMBH at $0.9$ and $1.1\,{\rm kpc}$ from the
Galactic center.}
\end{figure}

Figure \ref{fig:Surface-density-profiles} shows the surface density
profiles of our best fit models for M10 along with the observationally
derived profile, and in Tables \ref{tab:Evolution-Best-fit-Nobh}
and \ref{tab:Evolution-Best-fit-BH} we list their initial and final
cluster parameters. 

\begin{table}
\begin{tabular}{|c|c|c|}
\hline 
 & $t=0$ & $t=12\,{\rm Gyr}$\tabularnewline
\hline 
\hline 
$N$ & $5.0\times10^{5}$ $\left(5.7\times10^{5}\right)$ & $1.8\times10^{5}$ $\left(1.5\times10^{5}\right)$\tabularnewline
\hline 
$M_{c}$~$\left[{\rm M_{\odot}}\right]$ & $3.2\times10^{5}$ $\left(3.6\times10^{5}\right)$ & $7.8\times10^{4}$ $\left(6.9\times10^{4}\right)$\tabularnewline
\hline 
$r_{h}\,{\rm \left[pc\right]}$ & $3.85$ $\left(3.80\right)$ & $4.6$ $\left(3.90\right)$\tabularnewline
\hline 
$t_{rh}\,\left[{\rm Gyr}\right]$ & $1.6$ $\left(1.7\right)$ & $4.3$ $\left(4.0\right)$\tabularnewline
\hline 
$r_{tide}\,\left[{\rm pc}\right]$ & $25.8$~$\left(23.5\right)$ & $16.0$~$\left(13.6\right)$\tabularnewline
\hline 
\end{tabular}

\caption{\label{tab:Evolution-Best-fit-Nobh}Evolution of the characteristics
of our best-fit models without IMBH. Shown are the parameters for
the model with $R_{G}=1.1\,{\rm kpc}$ and, in parentheses, for the
model with $R_{G}=0.9\,{\rm kpc}$. Here, $r_{h}$ is the half-mass
radius, $r_{t}$ the tidal, or Jacobi, radius, and $M_{c}$ the total
cluster mass.}
\end{table}
\begin{table}
\begin{tabular}{|c|c|c|}
\hline 
 & $t=0$ & $t=12\,{\rm Gyr}$\tabularnewline
\hline 
\hline 
$N$ & $5.7\times10^{5}$ & $1.7\times10^{5}$\tabularnewline
\hline 
$M_{c}$~$\left[{\rm M_{\odot}}\right]$ & $3.6\times10^{5}$ & $7.7\times10^{4}$ \tabularnewline
\hline 
$r_{h}\,{\rm \left[pc\right]}$ & $3.4$  & $4.1$ \tabularnewline
\hline 
$t_{rh}\,\left[{\rm Gyr}\right]$ & $1.5$ & $4.1$\tabularnewline
\hline 
$r_{tide}\,\left[{\rm pc}\right]$ & $23.5$ & $14.0$\tabularnewline
\hline 
\end{tabular}

\caption{\label{tab:Evolution-Best-fit-BH} Evolution of the characteristics
of our best-fit models with IMBH. Shown are the parameters for the
model with a final $M_{BH}=480\,{\rm M_{\odot}}$, and $R_{G}=0.9\,{\rm kpc}$.}
\end{table}

\begin{figure}
\includegraphics[clip,width=0.9\columnwidth]{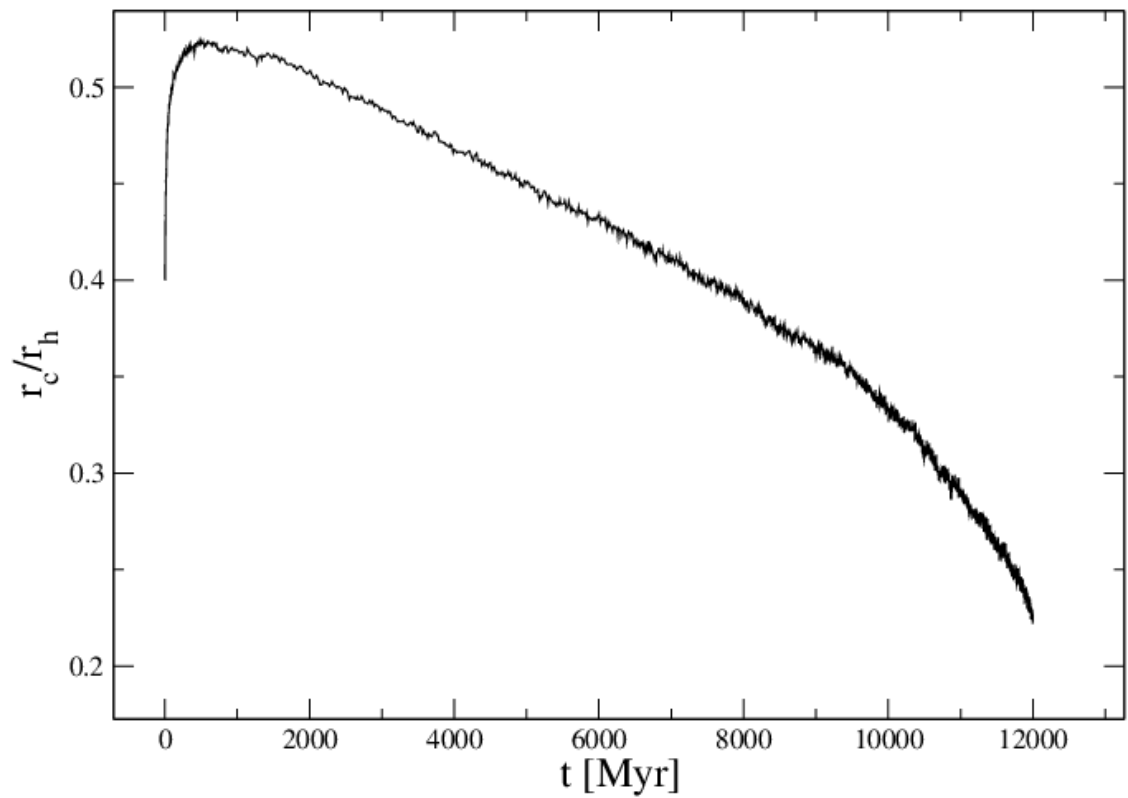}

\caption{\label{fig:Core-to-half-mass-radius}Core-to-half-mass radius over
time for the best-fit model without IMBH at $R_{G}=0.9\,{\rm kpc}$. }
\end{figure}

As can be seen, we are able to find models that match the surface
density profile out to approximately $400"$, or $8.5\,{\rm pc}$,
for both cases, with and without IMBH. Outside of this radius, the
observed profile flattens, which could be attributed to background
stars or escaping stars that remain still close to the cluster. This
could be due to the significant non-sphericity of the combined cluster
and Galactic potential \citep{2000MNRAS.318..753F,2010MNRAS.407.2241K},
which cannot be modeled with the simple tidal cut-off prescription
used in our code \citep[see][]{2010ApJ...719..915C}. Given the rather
large core to half-light ratio of M10 of $\approx0.4$ \citep{1996AJ....112.1487H},
it is not surprising that the models without IMBH are still in the
core contracting phase. This is also shown in Figure \ref{fig:Core-to-half-mass-radius},
where the evolution of the corresponding core-to-half-mass ratio is
shown. It should be noted here that the core radius in this figure
is calculated from both stars and stellar remnants, and, due to the
strong concentration of the dark remnants towards the cluster center,
has a significantly smaller value at the end of the simulation.

From Figure \ref{fig:Surface-density-profiles} we also see that M10
can only harbor an IMBH with at most $580\,{\rm M_{\odot}}$. Given
the low final mass of the clusters of $\approx8\times10^{4}\,{\rm M_{\odot}}$,
this corresponds to a BH-to-cluster mass ratio of $0.75\%$, much
larger than what we would expect from the extrapolated $M_{BH}-\sigma$
relation \citep{FerrareseMerritt,2001bhbg.conf..246V,1995ARA&A..33..581K}.
However, the main reason for the large value is that the cluster lost
$80\%$ of its initial mass, while initially the BH-to-cluster mass
ratio was $0.15\%$, in good agreement with the $M_{BH}-\sigma$ relation.

\subsection{Surface Brightness Profile}

An alternative method to detect imprints of an IMBH is to construct
the SBP from integrated light measurements. The advantages and disadvantages
of integrated light profiles with respect to star count profiles have
been discussed by \citet{2006AJ....132..447N} and \citet{2011ApJ...743...52N}.
In short, star count profiles suffer from crowding, whereby faint
stars cannot be detected in the vicinity of bright stars affecting
the overall profile shape. Therefore, star count profiles are usually
limited to fewer, brighter stars with high completeness, which in
turn, however, makes the profiles noisier. In contrast, integrated
light profiles contain the contribution of all stars, but a few very
bright giants can contribute disproportionately to it, resulting in
larger noise levels. The brightest stars are, therefore, removed before
the profile is calculated. In real images this subtraction cannot
be done cleanly and there is always a hard-to-quantify contribution
from the wings of the point-spread-function of subtracted giants \citep{2006AJ....132..447N}.
An accurate comparison to theoretical models, thus, would require
producing synthetic images from the models and repeating exactly the
same procedure that has been done to derive the observed profile,
as has been carried out in \citet{2011ApJ...743...52N}. Here, we
chose instead a simpler procedure by removing all stars above a certain
cut-off magnitude with a value chosen as a compromise between low
noise levels and minimal changes in the profile shape. 

\begin{figure}
\includegraphics[clip,width=0.95\columnwidth]{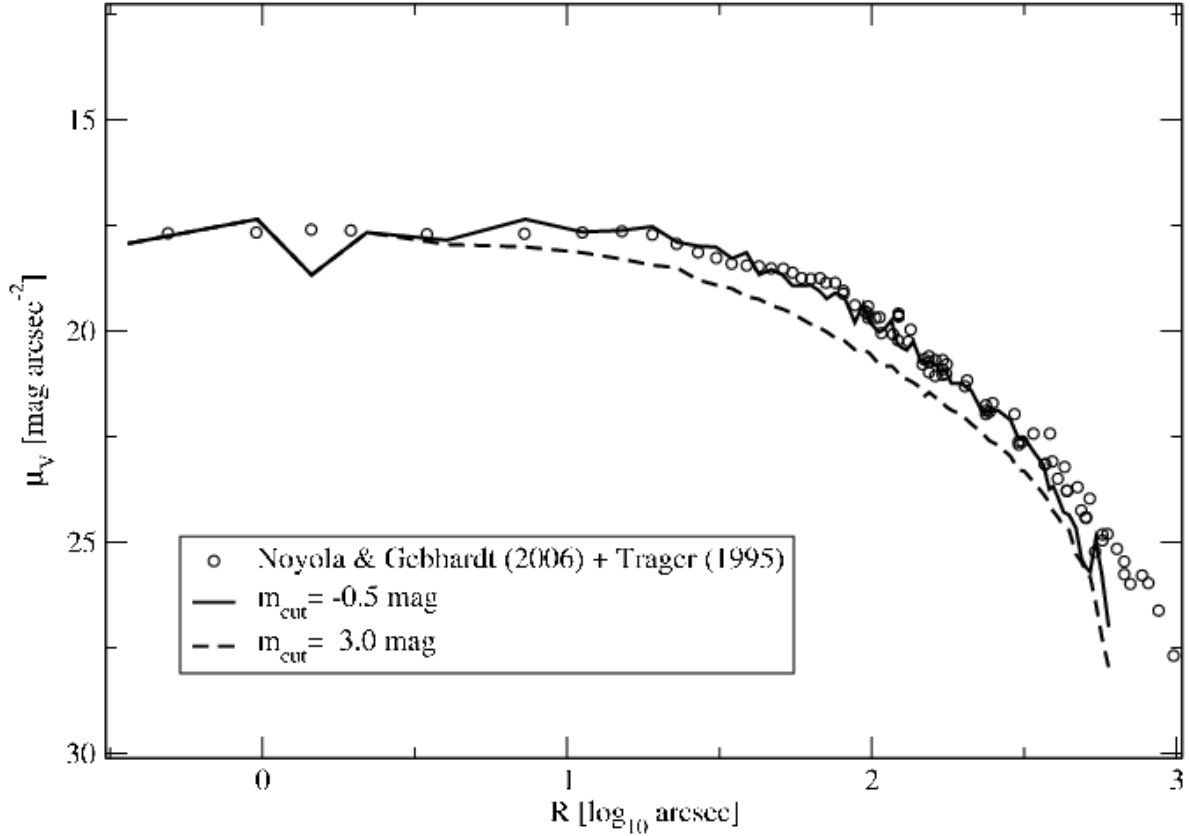}

\caption{\label{tab:Surface-brightness-profile}Surface brightness profile
for the best fit model without IMBH. Shown is the combined observed
profile (open circles) from \citet{2006AJ....132..447N} and \citet{1995AJ....109..218T}
(inside and outside $76\,{\rm arcsec}$ respectively), as well as
the profile of our best fit model with $R_{G}=0.9\,{\rm kpc}$ and
two absolute cut-off V-band magnitudes, $3\,{\rm mag}$ (dashed line)
and $-0.5\,{\rm mag}$ (solid line), for the masking of bright giants.
The model SBP follows the observed profile well and both, observed
SBP and SDP, are, thus, consistent with each other. The influence
of the cut-off magnitude on the SBP is, with $\approx1\,{\rm mag}$
difference in the profiles, significant and much larger than for the
models in \citet{2012ApJ...750...31U} ($<0.2\,{\rm mag}$ for the
same cut-off values). }

\end{figure}
 Figure \ref{tab:Surface-brightness-profile} shows the SBP of the
best-fit M10 model from the previous section with $R_{G}=0.9\,{\rm kpc}$,
applying two cut-off magnitudes to mask bright giants. As can be seen,
the model SBP follows the observed profile well, and, therefore, the
observed surface brightness and the star count data, despite being
derived from rather different observations, and possibly constructed
relative to different cluster centers, appear to be consistent with
each other. However, in contrast to our previous models for NGC~5694,
the influence of the absolute cut-off magnitude is profound, and giants
up to an absolute magnitude of $-0.5\,{\rm mag}$ contribute significantly
to the total cluster light throughout. The reason for this could be
related to the strong tidal stripping that led, together with the
mass lost from stellar evolution, to about $80\%$ mass loss for our
model clusters at the end of the simulations. Since preferentially
low-mass stars are removed through this process, the low-mass end
of the mass function flattens \citep[e.g.,][]{2003MNRAS.340..227B},
and the brighter, more massive giants contribute more to the total
cluster light. Clearly, a simple cut-off prescription in absolute
magnitude is, in this case, not sufficient to obtain a smooth profile
and match the observed profile. This limits the extent to which we
can put constraints on the IMBH mass.

\begin{figure}
\includegraphics[clip,width=0.95\columnwidth]{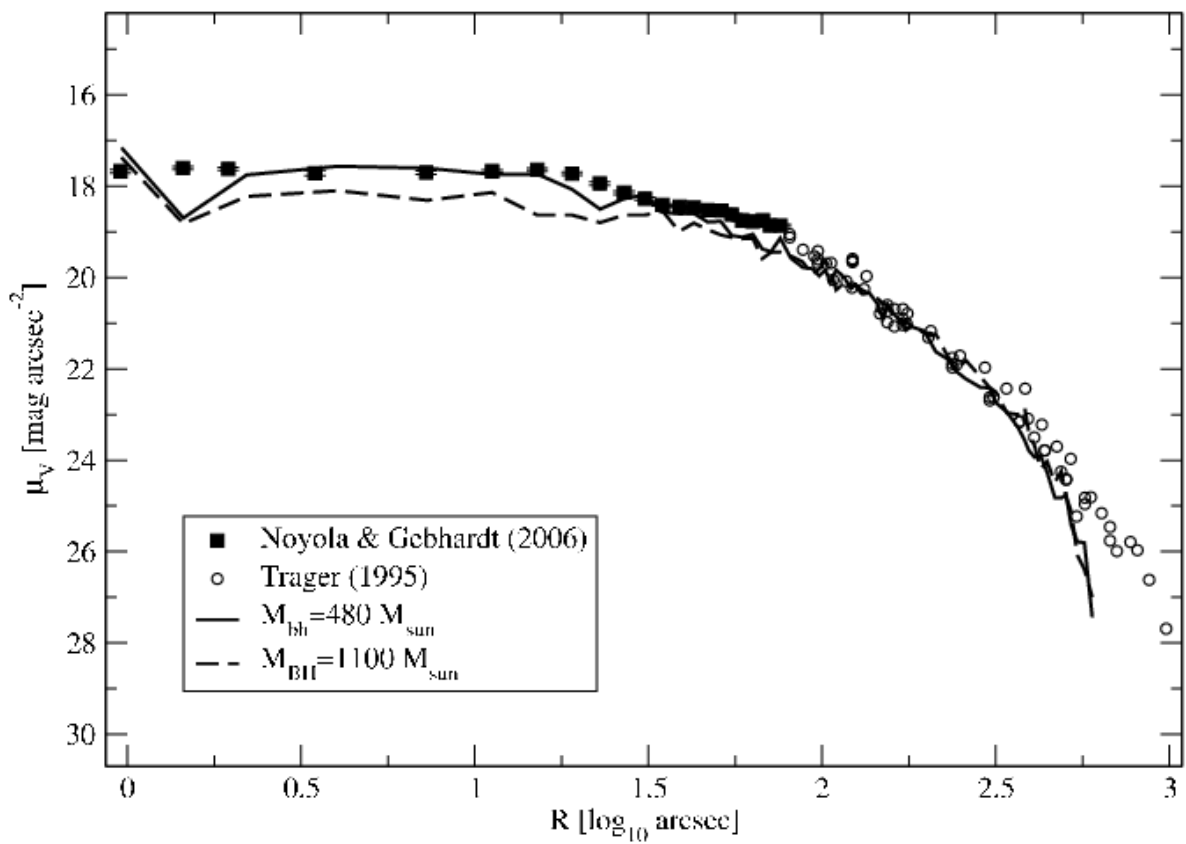}

\caption{\label{fig:Surface-brightness-profiles-bh}Surface brightness profiles
of models with an IMBH. Shown here are the observed profiles of \citet{2006AJ....132..447N}
(filled squares) and \citet{1995AJ....109..218T} (open circles) and
two model profiles with IMBH masses of $480\,{\rm M_{\odot}}$ (solid
line) and $1100\,{\rm M_{\odot}}$ (dashed lines). The $480\,{\rm M_{\odot}}$
IMBH model is the same as in Fig. \ref{fig:Surface-density-profiles}.
Due to the noise in the profiles, caused by very bright giants, it
is difficult to discern models with a difference in IMBH mass of less
than $500\,{\rm M_{\odot}}$.}
\end{figure}
Figure \ref{fig:Surface-brightness-profiles-bh} shows our results
for models with IMBH. Both models have the same initial conditions
as in Figure \ref{tab:Surface-brightness-profile} and only differed
in IMBH mass. As can be seen, the model with $M_{BH}=480\,{\rm M_{\odot}}$,
which is the same as the one in the previous section, fits the observed
profile reasonably well. However, because of the noise in the profile,
only a model with $M_{BH}=1100\,{\rm M_{\odot}}$ can be clearly ruled
out, while models with smaller $M_{BH}$ are harder to distinguish.
A smoother profile is generally more desirable as it allows stronger
constraints to be placed on the mass of a hypothetical central IMBH.
More sophisticated and elaborate techniques, as, e.g., presented in
\citet{2011ApJ...743...52N}, are necessary for this purpose.

\subsection{Mass Segregation}

A more recently proposed diagnostic for the presence of an IMBH in
a GC \citep{2008ApJ...686..303G}, is based on the tendency for mass
segregation to be suppressed by strong interactions in the vicinity
of a central IMBH. The signature is, however, not unique, as binary
interactions lead to a similar decrease in the average mass provided
the binaries are sufficiently numerous and hard. \citet{2010ApJ...713..194B}
determined through deep photometry with the ACS camera on\emph{ HST}
the mean mass profile for M10, which is defined as \citep{2009ApJ...699.1511P}
\[
\Delta m(r)=\left\langle m\right\rangle (r)-\left\langle m\right\rangle (r_{h})
\]
where $\left\langle m\right\rangle (r)$ is the mean stellar mass
at distance $r$ from the cluster center, and $r_{h}$ the half-mass
radius. They found that both a cluster with an IMBH, or a cluster
without IMBH but with a binary fraction $>5\%$, can reproduce the
observed profile. In an effort to draw firmer conclusions as to whether
an IMBH is required to explain the mean mass profile in M10, \citet{2011ApJ...743...11D}
compared the observed radial dependence of the binary fraction with
the corresponding results of the simulations in \citet{2010ApJ...713..194B}
and found that the observed fractions are always larger than in the
models with a global binary fraction of 5\%. They conclude that the
binary fraction is large enough to suppress mass segregation in the
center of M10 an an IMBH is not needed to explain the observations.

The mass segregation signature, however, develops only after the cluster
has had enough time to relax, typically $\sim5\, t_{rh}$ \citep{2008ApJ...686..303G}.
From Tables \ref{tab:Evolution-Best-fit-Nobh} and \ref{tab:Evolution-Best-fit-BH}
we already see that the half-mass relaxation times in our models are
rather large and vary by a factor of more than two during the cluster
evolution, which makes it not so clear whether M10 is really old enough
for the mass segregation signature to be discernible. To address this
issue, we analyze our best fit models following the same procedure
used for the $N$-body models in \citet{2010ApJ...713..194B}, and
compare to the corresponding observations.

\begin{figure}
\begin{tabular}{cc}
\includegraphics[clip,width=0.45\columnwidth]{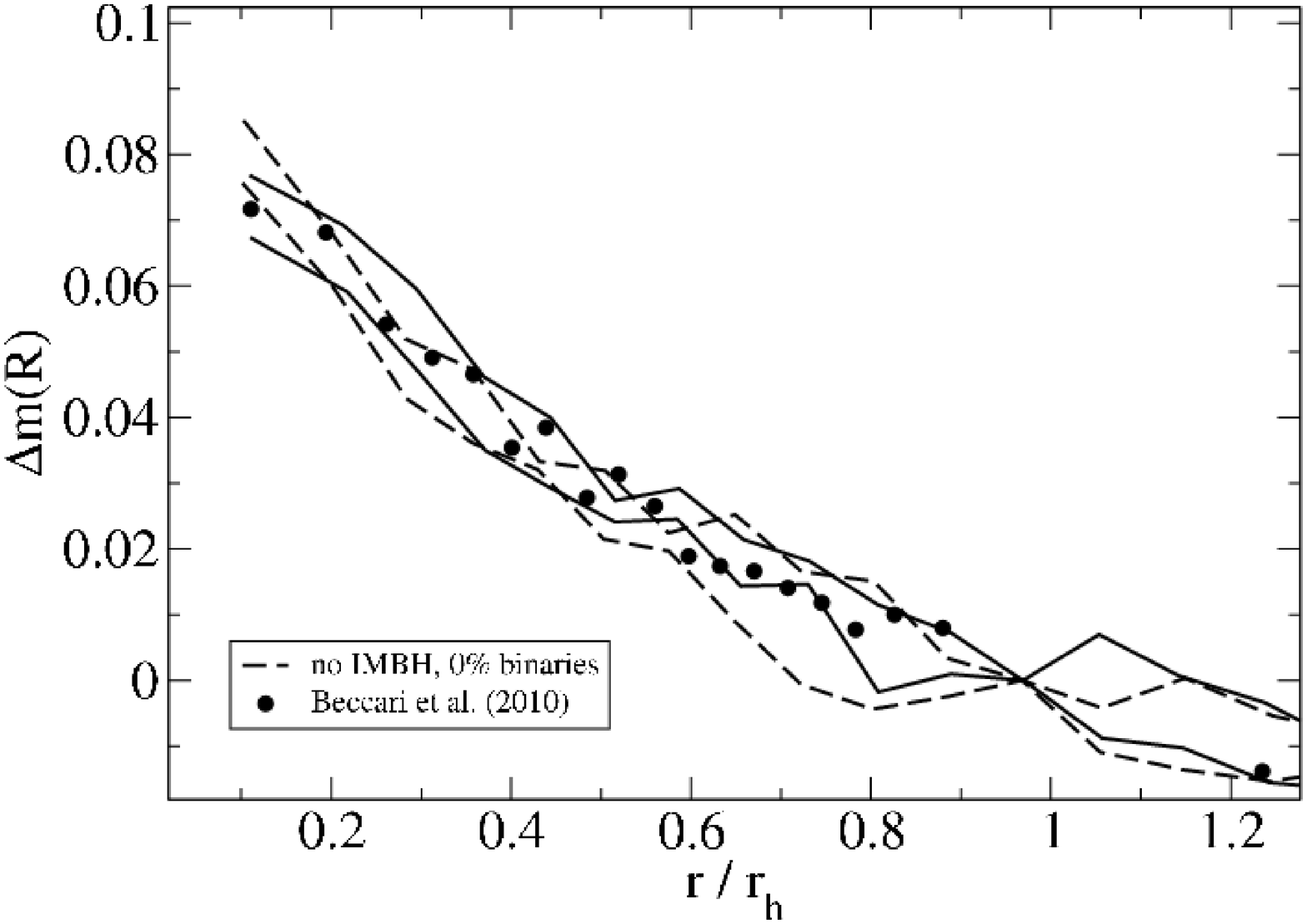} & \includegraphics[width=0.45\columnwidth]{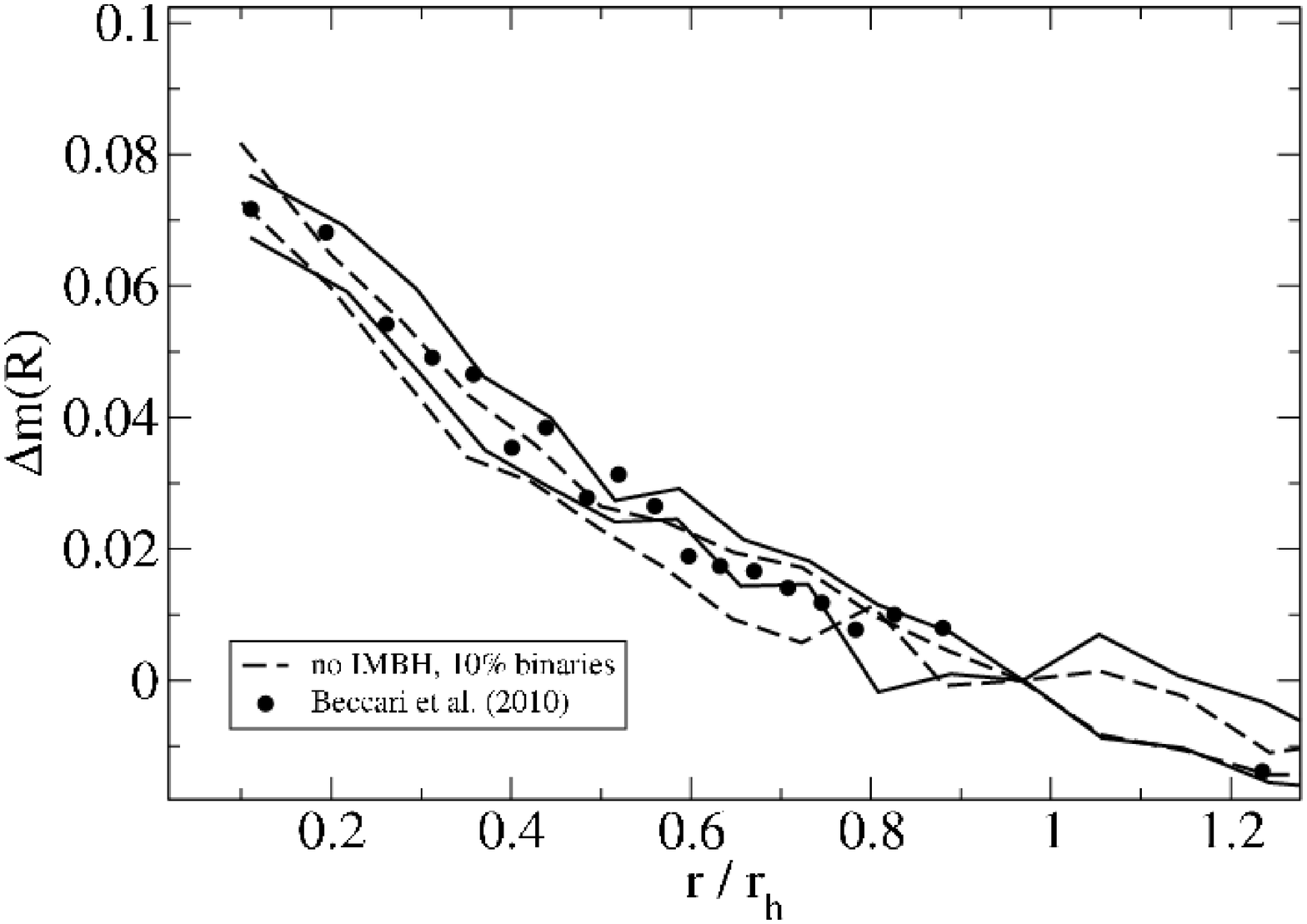}\tabularnewline
\end{tabular}

\caption{\label{fig:Average-mass-profile}Average mass profile for our best-fit
models with and without IMBH. Each final cluster state was projected
on the sky from 10 random directions, and shown are two of the resulting
profiles that roughly embrace all the others. Left: Results with and
without IMBH and no binaries. Right: Results with and without IMBH
and 10\% hard binaries. All models reproduce the average mass profile
very well, and there are only negligible differences between them.
The reason is most likely the young dynamical age of M10, as only
little more than two half-mass relaxation times have elapsed.}
\end{figure}
\begin{figure}
\includegraphics[clip,width=0.95\columnwidth]{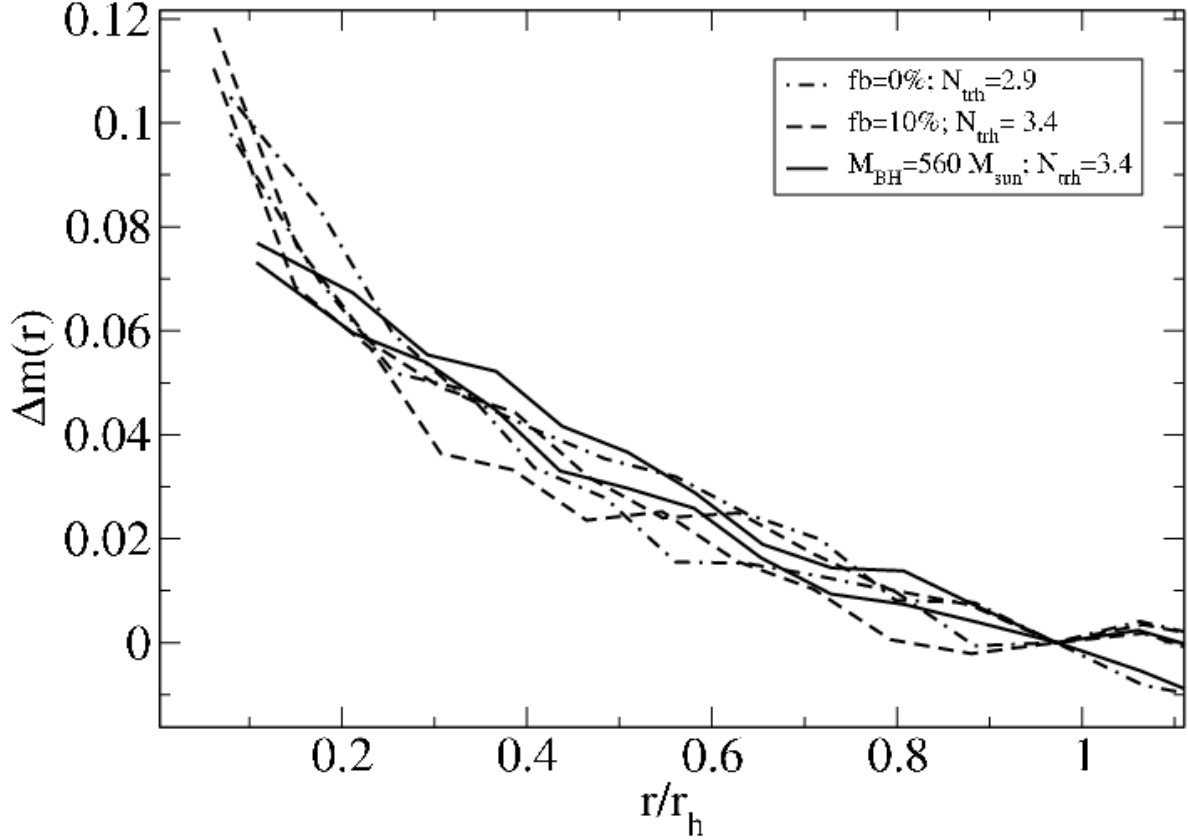}

\caption{\label{fig:Average-mass-profile-evolved}Mean mass profile for more
evolved clusters. All models have initial conditions as in Table \ref{tab:Evolution-Best-fit-Nobh}
and \ref{tab:Evolution-Best-fit-BH} but started more concentrated,
making them dynamically more evolved at $12\,{\rm Gyr}$. While there
is little change compared to Fig. \ref{fig:Average-mass-profile}
for the model with IMBH, the models without IMBH show a significantly
increased level of mass segregation in the center.}
\end{figure}

 Figure \ref{fig:Average-mass-profile} shows the mean mass profile
of M10 and our best fit cluster models with and without IMBH, including
models with 10\% and 0\% binaries. As can be seen, all model profiles
match the data in \citet{2010ApJ...713..194B} very closely, which
is somewhat surprising given that we would expect the model with no
binaries and no IMBH to be more strongly segregated than the others.
The differences, instead, are rather marginal, which could indeed
indicate that M10 is still dynamically young. When we calculate the
dynamical age as the number of elapsed half-mass relaxation times,
$N_{t_{rh}}$, defined by \citet{2007MNRAS.379...93H} as
\begin{equation}
N_{t_{rh}}=\int_{0}^{\tau}\frac{dt}{t_{rh}(t)}\label{eq:N_trh}
\end{equation}
where $\tau$ is the cluster age, we obtain $N_{t_{rh}}=2.6,\,2.4,\,2.8$
for the single star, binary and IMBH cluster models, respectively.
Thus, compared, e.g., to NGC~5694 \citep{2012ApJ...750...31U}, M10
appears to be dynamically only half as old. As can be seen in Figures
2 and 3 in \citet{2008ApJ...686..303G}, at these young ages the innermost
points, $\Delta m(0)$, still overlap significantly between the different
models, making it difficult, if not impossible, to discern between
clusters with and without IMBH. In order to illustrate this point
further, we also calculated $\Delta m(r)$ for dynamically more evolved
clusters, with a larger initial concentration ($W_{0}=7$ instead
of $W_{0}=5.6$) but otherwise the same initial conditions as our
best fit clusters. As Figure \ref{fig:Average-mass-profile-evolved}
 shows, the clusters without IMBH now have a clearly larger average
stellar mass at the center, while the model with IMBH has an average
mass profile that is very similar to the one in Figure \ref{fig:Average-mass-profile}.
This clearly confirms that our best fit models without IMBH have not
yet reached their fully relaxed state, and the reason for the lower
level of mass segregation in M10 could simply be a consequence of
its young dynamical age, rather than the presence of a IMBH in its
center.

\subsection{Velocity Dispersion}

One of the great advantages of modeling the evolution of clusters
on a star-by-star basis with the full number of stars is that the
resulting models allow us to make detailed predictions for observations.
This is especially important for the detection of the velocity dispersion
signature of IMBHs, as for the expected IMBH masses, the radial extent
of the inner cusp is small and does not contain many objects \citep{2004ApJ...613.1133B},
making the interpretation of the data very challenging. The problem
becomes even more severe considering that, through mass segregation,
the center of the cluster is dominated by dark remnants, further reducing
the number of bright stars available for velocity measurements in
the cusp. Here we determine to what extent we can expect to detect
an increase in velocity dispersion for the innermost bright stars
in the core of M10.

\begin{figure}
\begin{tabular}{cc}
\includegraphics[clip,width=0.45\columnwidth]{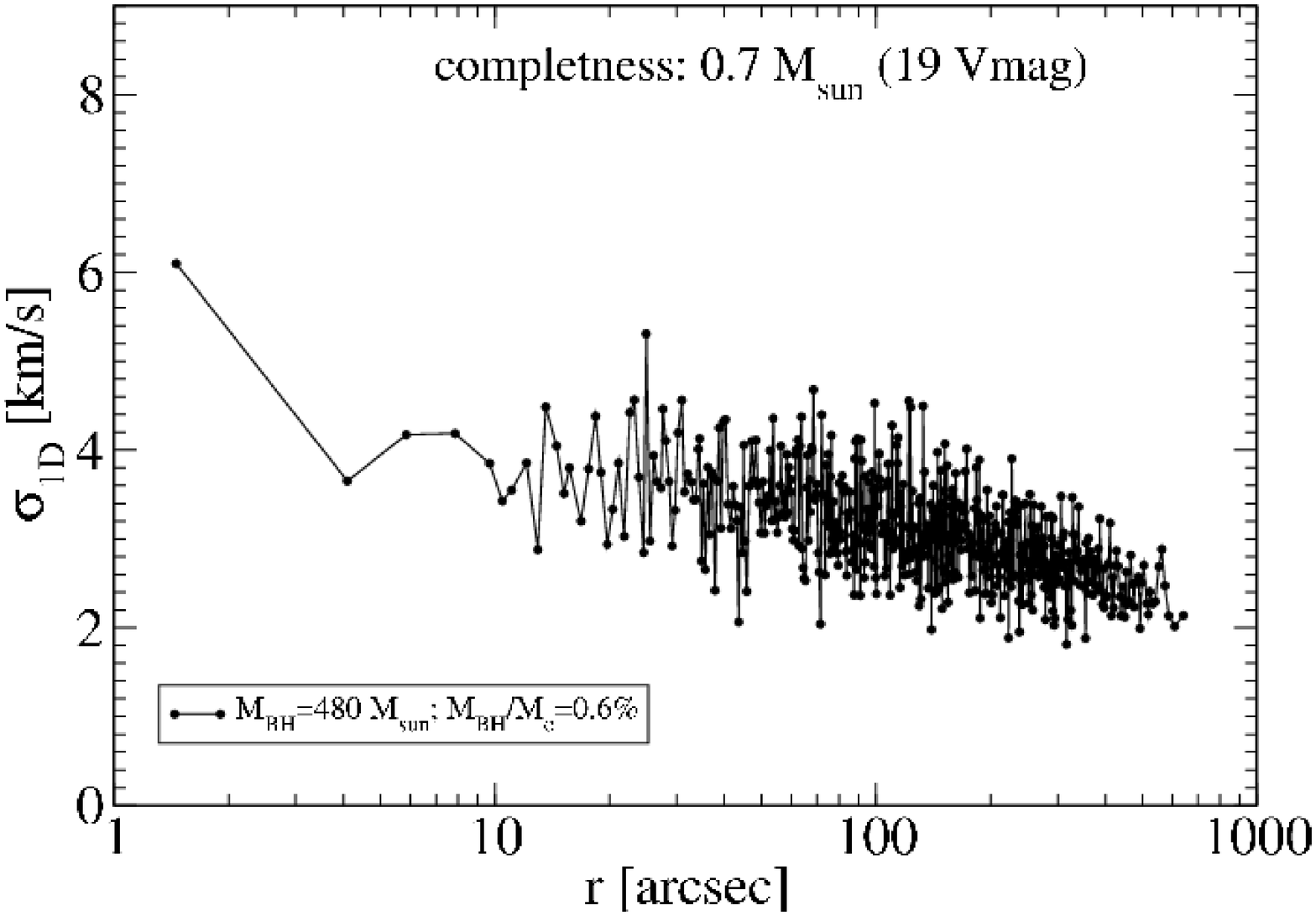} & \includegraphics[clip,width=0.45\columnwidth]{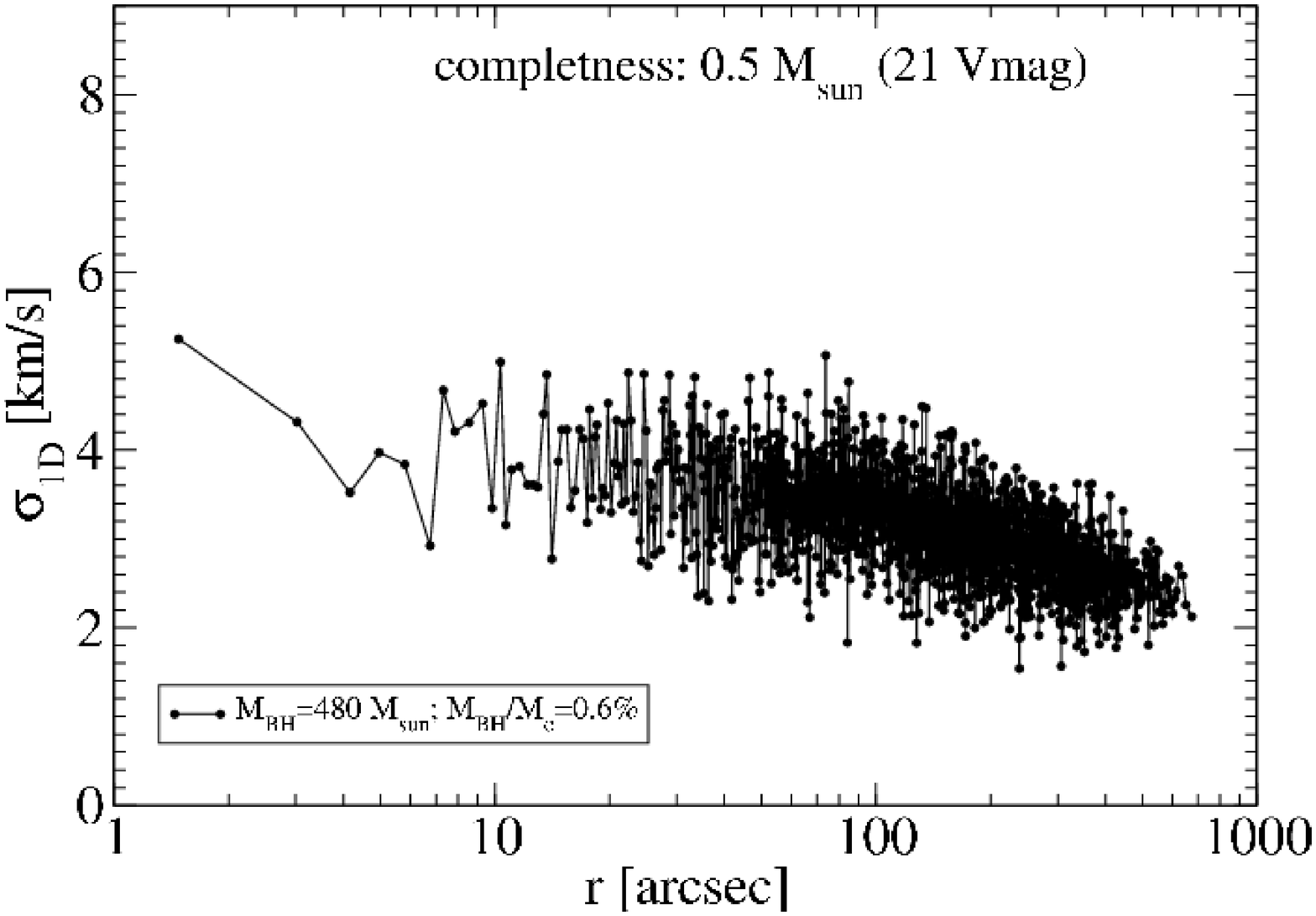}\tabularnewline
\includegraphics[clip,width=0.45\columnwidth]{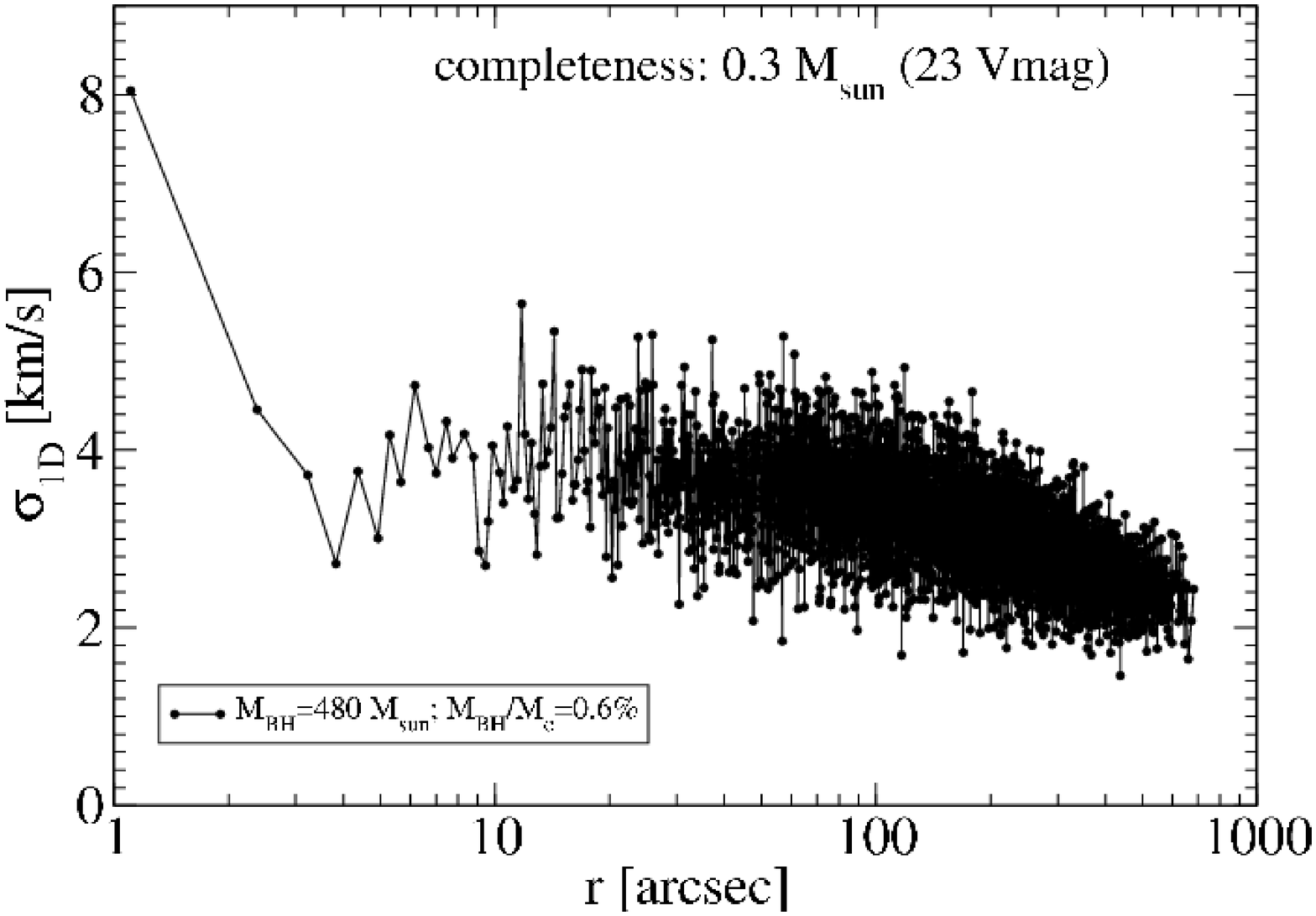} & \tabularnewline
\end{tabular}

\caption{\label{fig:One-dimensional-velocity-dispers}Projected one-dimensional
velocity dispersion profile of our best fit model with $M_{BH}=480\,{\rm M_{\odot}}$,
for various completeness levels. Each point represents the velocity
dispersion of 20 stars. The scatter of the values is of the order
of the Poisson error, and is about $1\,{\rm km\, s^{-1}}$. A cusp
with an extension of $3\,\arcsec$ is visible, but contains only $20-40$
stars. A significant increase in velocity dispersion is only detectable
if also low-mass stars, with masses down to $0.3\,{\rm M_{\odot}}$,
are considered. }
\end{figure}
 In Figure \ref{fig:One-dimensional-velocity-dispers} we show the
projected one-dimensional velocity profile of our model with $M_{{\rm BH}}=480\,{\rm M_{\odot}}$
including only bright main-sequence stars and giants above certain
brightness cut-offs. Each point in this profile represents the velocity
dispersion of 20 neighboring stars.

As can be seen, the velocity dispersion cusp extends out to $\approx3\,\arcsec$
and contains in each case $20-40$ bright stars. Furthermore, the
velocity dispersion of the innermost point is only about 50\% larger
than the average in the cluster core, even when the completeness limit
is as low as $0.5\,{\rm M_{\odot}}$. Given that the statistical fluctuation
of each point, as determined from 10 random projections of the cluster
on the sky, is $\approx2\,{\rm km\, s^{-1}}$, this increase is barely
significant. Only for a limit of $0.3\,{\rm M_{\odot}}$ does the
velocity dispersion of the innermost point become significantly larger,
and, with twice the value, can be clearly distinguished from the core
velocity dispersion. The reason for the increase in velocity dispersion
with decreasing brightness cut-off is that, as more stars are included,
the first 20 stars sample a region closer to the IMBH, where velocities
are larger.

From this we can see that detecting an IMBH with only about $500\,{\rm M_{\odot}}$
in M10 based on velocity dispersion measurements alone is extremely
challenging and would require very deep observations with completeness
limits down to $0.3\,{\rm M_{\odot}}$. This is despite the fact that
the radial extent of the cusp, out to about $3\,\arcsec$, is large
enough to be easily resolved. The sparseness of bright stars in the
central region prevents sampling the velocity dispersion close enough
to the IMBH to detect a significantly increase above the core velocity
dispersion.

\section{Summary and Conclusions}

We carried out Monte Carlo simulations to constrain IMBHs at the centers
of observed GCs considering a variety of observational diagnostics.
In contrast to our earlier study \citep{2012ApJ...750...31U}, we
focused on a cluster, M10, that undergoes significant tidal stripping,
resulting in final cluster masses of only about 20\% of the initial. 

From a comparison to a detailed star count profile we were able to put an upper limit
on the mass of a hypothetical IMBH in M10. We find that the maximum
IMBH mass that results in a SDP that still fits the observed profile
is $\approx600\,{\rm M_{\odot}}$. This IMBH mass also also leads
to a SBP that is compatible with observations by \citet{2006AJ....132..447N},
although due to the large radial fluctuations caused by very bright
giants, the SBP is less constraining. This is in contrast to our earlier
investigation of the cluster NGC~5694 \citep{2012ApJ...750...31U}
where the SBP was much less noisy. The reason is mostly likely that
the light in M10 is dominated by bright giants as we find that the
model profile shape is rather sensitive to the chosen cut-off brightness.
This could be a direct consequence of the preferential loss of low-mass
stars through tidal stripping, which causes the low-mass end of the
mass function to flatten, and the brighter, more massive giants contribute
more to the total cluster light. The more detailed analysis in \citet{2011ApJ...743...52N}
might lead to a much smoother profile, and, therefore, could have
the potential to obtain stronger constraints for an IMBH in M10 than
star count data, as integrated light measurements generally include
the contribution of many more stars.

In addition to light and star count profiles, we also considered the
amount of mass segregation in M10 as an indicator of an IMBH. Our
main finding is that M10 is not dynamically relaxed enough for the
IMBH signature, a relative depression of the mean stellar mass at
the cluster center, to be detectable. Indeed, when calculating the
number of elapsed half-mass relaxation times (Equation \ref{eq:N_trh}),
M10 is dynamically only half as old as NGC~5694 and still in its
core contraction phase. Similar to the case of NGC~5694 in \citet{2012ApJ...750...31U},
this demonstrates that simple dynamical age estimates based on the
current state of globular cluster are extremely uncertain. We also
showed that interactions of stellar binaries are not able to sustain
the large core of M10 ($r_{c}/r_{h}\approx0.4$; Harris 1996) even
with a binary fraction as large as 10\%.

Finally, we found that the velocity dispersion signature of an IMBH
in our best-fit M10 model is very challenging to detect. This is because
the Keplerian cusp is rather sparsely populated with bright stars
and main-sequence stars with masses down to $0.3\,{\rm M_{\odot}}$
have to be included in a future velocity measurement in order to detect
a significant increase for the innermost 20-40 stars. Given that $0.3\,{\rm M_{\odot}}$
stars have been detected with only a 50\% completeness in the core
of M10 by \citet{2010ApJ...713..194B}\emph{, }detecting such an increase\emph{
}remains difficult even when a second epoch of ACS data with a sufficiently
long time baseline becomes available.

\acknowledgements{}

We thank Barbara Lanzoni for providing us with the star count data
for M10 from HST archival data (GO-10775; PI: A. Sarajedini). S.U.
was supported by Hubble Theory Program HST-AR-11779, provided by NASA
through a grant from the Space Telescope Science Institute, which
is operated by the Association of Universities for Research in Astronomy,
Inc., under NASA contract NAS 5-26555. We also acknowledge support
from NASA ATP Grant NNX09AO36G at Northwestern University. This work
was started while the authors and Barbara Lanzoni were participants
of the program ``Formation and Evolution of Globular Clusters''
at KITP in Spring 2009. Our work at KITP was supported in part by
the National Science Foundation under Grant No. NSF PHY11-25915.

\end{document}